# Noise-like pulse laser source with ultrabroadband tunability and coherence-limited sub-structure

Xinyang Liu ,[1,2,*] Matias Koivurova[3] and Regina Gumenyuk[1]

[1]Laboratory of photonics, Tampere University, Korkeakoulunkatu 3, Tampere 33720, Finland

[2]Russell Centre for Advanced Lightwave Science, Shanghai Institute of Optics and Fine Mechanics and Hangzhou Institute of Optics and Fine Mechanics, Hangzhou 311421, China

[3]Center for Photonics Sciences, University of Eastern Finland, P.O. Box 111, FI-80101 Joensuu, Finland

## Abstract

High brightness and low coherence laser sources with wideband tunability are essential for many full-field imaging applications aiming for high contrast and speckle-free performance. However, this combination of parameters is challenging to achieve. The current solutions focus on decreasing spatial coherence or generation of time-varying speckle patterns, while suppression of temporal coherence typically compromises brightness. Here we demonstrate a wideband pulsed laser source with low temporal coherence and the absence of phase correlation between pulses as an alternative approach with simultaneous time and frequency diversity. The full-gain-spectrum of a Tm-doped fiber laser (1650 nm – 2000 nm) is operated in a tunable noise-like pulse regime, which by nature is composed of countless structured elementary events with uncorrelated phases randomly varying from bunch to bunch. The measured spectral widths range from 13.8 nm to 18.8 nm, while the average output power varies between 63.3 mW and 213 mW. Numerical simulations reveal that temporal coherence decreases significantly with increasing optical gain, dropping from near unity at low gain to approximately 0.2 at high gain. The startup dynamics of the noise-like pulse laser are experimentally studied using the dispersive Fourier transformation (DFT) method. Based on single-shot spectra and frequency-resolved optical gating traces, the coherence properties of the laser are further analyzed by calculating the mutual coherence function and cross-spectral density. The noise-like pulse laser exhibits a coherence time of approximately 100 fs and an average pulse-burst duration of about 40 ps in the high-gain regime.

## Introduction

Light sources are a fundamental component of any optical imaging system, as their physical properties directly determine image contrast, resolution, penetration depth, and the presence of artifacts[1-4]. Key characteristics include spectral bandwidth, temporal and spatial coherence, brightness (radiance), and emission stability. The spectral bandwidth influences axial resolution and coherence length, while spatial and temporal coherence govern interference effects and speckle formation. Brightness determines the achievable signal-to-noise ratio and imaging speed, particularly in low-light or nonlinear modalities. Consequently, selecting or engineering an appropriate light source requires balancing these parameters to meet the specific demands of a given imaging technique.

In terms of temporal (first-order coherence) and spatial coherence, as well as brightness characteristics, light sources can generally be classified into the four categories illustrated in Fig. 1. The narrow-linewidth DFB fiber laser features both high temporal and spatial coherence. The high temporal coherence results from its narrow linewidth, while the high spatial coherence is determined by the single-mode fiber waveguide. As all power is concentrated in a narrow bandwidth, this type of laser is typically characterized also by high brightness. The DFB fiber laser is particularly of use for optical communication[5] and optical sensing[6]. Among ultrashort pulsed lasers, the supercontinuum source is widely favored in optical imaging, particularly, in optical coherence tomography (OCT) applications[7]. Its distinctive characteristic is an ultrawide spectral band generated in a single-mode optical waveguide via interaction between a high-intensity laser beam and material nonlinearity. This results in short coherence time and high spatial coherence. However, as the supercontinuum source originates from an ultrashort pulse laser, its pulse-to-pulse phase coherence (second-order coherence) remains very high, and the brightness of some wavelength bands is severely degraded due to uneven energy distribution over the wide spectrum. Another type of laser used for optical imaging is a random laser. It suppresses coherence in both temporal and spatial domains, rooted in its peculiar scattering scheme during light generation. Low spatial coherence helps to avoid coherence artifacts in imaging applications[8], but at the same time compromises low brightness. An alternative option is a fluorescence source featuring both low temporal and spatial coherence, which originates from spontaneous emission. Low temporal and spatial coherence ensure the capturing of clear images, although suffering from lowered brightness. As a result, in general, light sources exhibit an inherent trade-off between coherence and brightness: high brightness is typically associated with high coherence, whereas low coherence is often accompanied by reduced brightness. Consequently, achieving an optimal balance between these parameters within a single light source remains highly challenging.

In optical imaging, first-order coherence plays the dominant role in determining interference effects and speckle formation. Therefore, in the last decade, ultrashort pulses have attracted a lot of interest in this field. Owing to their broad spectral bandwidth, ultrashort pulses exhibit a short temporal coherence length, which can significantly reduce coherence artifacts and speckle contrast compared to narrowband continuous-wave lasers. This limited first-order temporal coherence is particularly advantageous in techniques such as OCT, where axial sectioning relies on controlled interference over short path length differences. Nevertheless, ultrashort pulse lasers typically retain high spatial coherence, so residual speckles can still arise in scattering media when optical path differences remain within the coherence length. Second-order coherence effects are generally minor in linear imaging applications, as most ultrashort pulse lasers operate close to coherent-state photon statistics with weak intensity correlations. However, in nonlinear imaging modalities such as multiphoton microscopy, intensity-dependent processes can make residual second-order coherence properties more relevant, influencing signal fluctuations and contrast.

High second-order coherence is a typical characteristic of mode-locked lasers, arising from the fundamental nature of the technique in which multiple longitudinal modes oscillate with a fixed phase relationship. This phase locking results in a stable train of ultrashort pulses with high temporal coherence. Among various mode-locked laser platforms, fiber laser cavities are widely used for generating pulses with diverse temporal and spectral characteristics. Depending on the dispersion and nonlinear dynamics within the cavity, different pulse regimes can be obtained, including the $sech^2$-shaped conventional soliton[9], parabolic similariton[10], and higher-order Gaussian-shaped dissipative soliton[11]. In addition to their role as ultrafast pulse sources, these fiber lasers are also commonly used as seed lasers for supercontinuum generation. A particular mode-locking regime of interest is noise-like pulse (NLP) operation, in which the output pulse consists of a large number of randomly distributed sub-pulses[12-14]. Instead of a single well-defined pulse, an NLP fiber laser emits pulse bunches with picosecond-scale envelope duration that contain numerous femtosecond sub-pulses. These pulse bunches repeat at the fundamental mode-locking repetition rate determined by the cavity length.

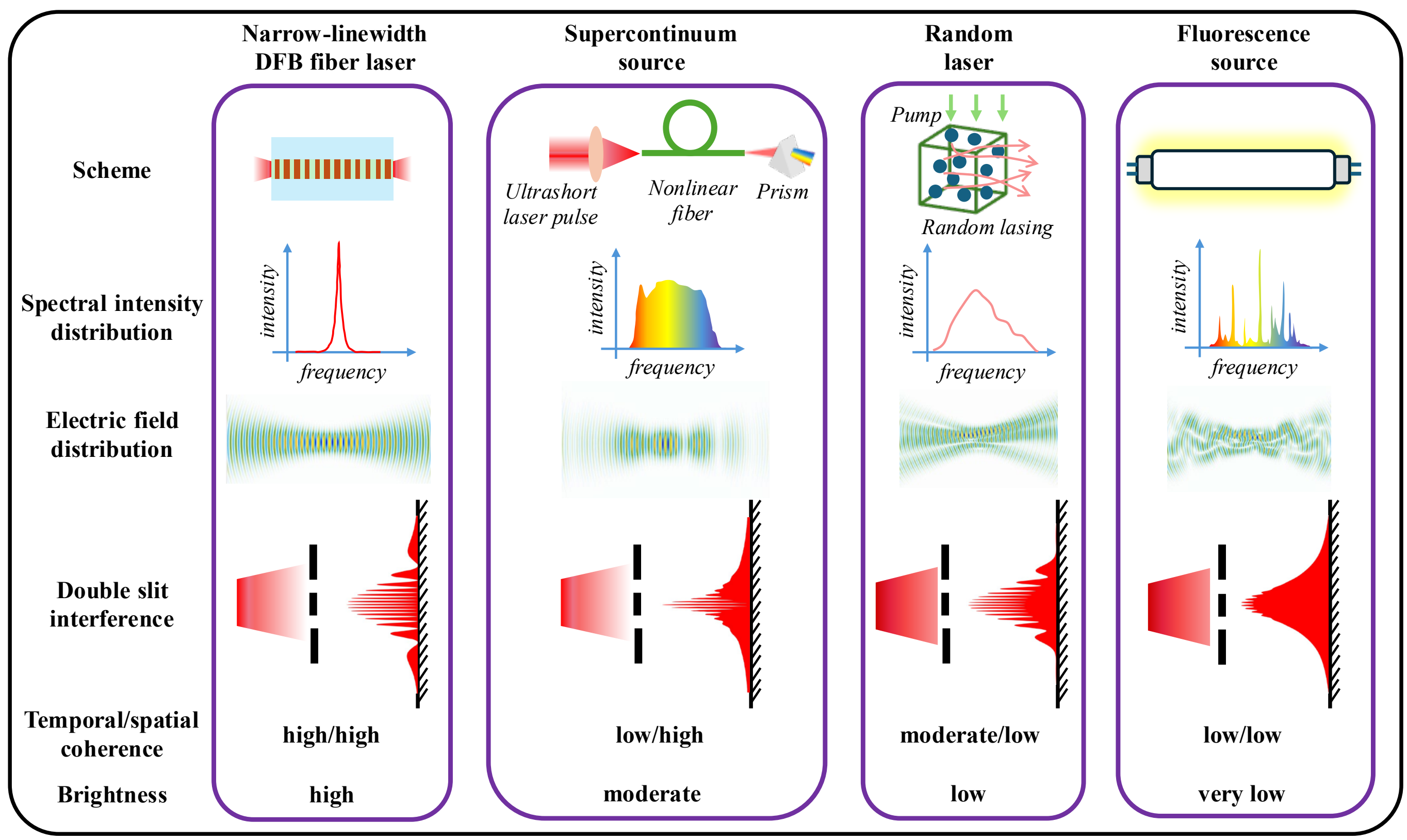

Fig. 1 Representative light sources with different coherence and brightness properties.

Despite the periodic emission of pulse envelopes, the internal structure of NLPs is intrinsically stochastic due to the random distribution and interaction of the constituent sub-pulses. Consequently, unlike conventional mode-locked regimes where a stable phase relationship between modes ensures high coherence, NLP operation is expected to exhibit reduced coherence. The random temporal and spectral fluctuations within the pulse envelope can lead to significant intensity fluctuations and therefore a degradation of the second-order coherence. The only attempt to directly measure the coherence of NLP was unable to give a quantitative result, concluding qualitatively that the temporal coherence was low[15].

Beyond brightness and coherence properties, the operating wavelength of an imaging light source is a key factor governing light propagation in biological tissue. In particular, significant attention has been devoted to wavelength regions that correspond to the so-called biological optical windows, where light experiences reduced absorption and scattering in biological tissues. These spectral windows enable deeper light penetration and therefore facilitate biomedical imaging and therapeutic applications. While the first and second biological windows, located approximately around 650–950 nm and 1000–1350 nm, respectively, have been widely explored, increasing interest has recently shifted toward the third biological window in the 1550–1870 nm range[16-18]. Light in this spectral region experiences reduced photon scattering in tissue, which enables improved imaging contrast and deeper penetration depth[18]. Consequently, fiber laser sources operating near 1.7–2.0 μm, including those supporting NLP regimes, are attracting growing attention for biomedical photonics and related applications.

In this work, we demonstrate an NLP laser with ultrabroadband tunability, covering the operating wavelength range from 1650 nm to 2000 nm. By employing an intracavity acousto-optical tunable filter (AOTF) as a wideband spectral filter, the full gain spectrum of Tm-doped fiber is explored for NLP operation. The laser output powers range from 63.3 mW to 213 mW with 3-dB spectral widths varying from 13.8 nm to 18.8 nm. The duration of the central spike is ~800 fs, and average pulse burst length is ~40 ps. The investigation of the transition dynamics between continuous-wave (CW), mode-locking and NLP states are performed both experimentally by exploiting DFT analysis and by numerical simulation, revealing the difference of the build-up process at different gain values. Comprehensive temporal coherence characterization of the output pulses exposes their quasi-stationary nature, such that the correlations follow the Schell-model, enabling us to unambiguously estimate the coherence time. The pulses have a coherence time of ~100 fs (coherence length of ~30 μm), and an average pulse length of about 40 ps.

## Results

### Noise-like pulse laser with ultrawideband tunability

Fig. 2 (a) depicts the experimental setup of the NLP fiber laser based on the ring cavity architecture. The laser cavity is composed of total 5.5 m anomalous-dispersion fiber with AOTF and three waveplates inserted in the free space part. Two quarter waveplates (λ/4) and one half waveplate (λ/2) control the polarization rotation process and enable the nonlinear polarization rotation (NPR) mode-locking mechanism. The fundamental repetition rate is 35 MHz. Two collimating lenses (L1 and L2) are used for free space laser beam collimation and coupling the laser beam back into the cavity. One polarization-independent isolator is used to keep unidirectional light propagation. To provide enough optical gain for NLP operation, two 1550 nm master oscillator power amplifiers with output powers of 1.7 W and 1.5 W are employed to excite the 0.7 m Tm-doped fiber (TDF) from both ends. The output coupler has a 20 % output power coupling ratio. The AOTF has a transmission bandwidth of 12 nm and a Gaussian shape transmission profile. Additionally, the AOTF also provides a frequency shifting effect with a frequency shift value determined by the driven RF frequency, adding another mode-locking regime that can help suppress CW and low intensity light[19]. The loss introduced by AOTF is negligible, while the fiber coupling loss after AOTF is around 30%. NLP operation can be obtained at the highest pump powers for all wavelengths covering the wavelength band from 1650 nm to 2000 nm. The center wavelength can be tuned from 1712 nm to 1957 nm through merely changing the driving frequency of the AOTF from 31.2 MHz to 27 MHz, as shown in Fig. 2(b, e-i), confirming the simplicity of the AOTF as a wavelength-tuning component. At the shortest wavelength, the sharp edge of the spectrum at around 1650 nm is caused by the cutting wavelength of the wavelength division multiplexer . At low pump power, and by setting the waveplates to appropriate angles, conventional soliton operation can be obtained. However, NLP operation cannot be obtained by only increasing pump power. Instead, it requires rotating waveplates while keeping the pump power at a high level.

Fig. 2(b) shows the obtained output powers and 3-dB spectral widths at different wavelengths. The output powers at different wavelengths range from 63.3 mW to 213 mW, whereas the 3-dB widths vary from 13.8 nm to 18.8 nm, following the same pattern. The variation of output powers follows the shape of the gain spectrum (see Fig. 2(d)), which is all limited by the available pump power. The increase of spectral widths towards the center of the Tm-doped gain is observed as a typical feature of noise-like pulse operation at the condition of stronger pumping [20]. Fig. 2(c) illustrates the characteristic intensity autocorrelation trace at 1712 nm with a narrow spike (~800 fs) positioning on a pedestal with tens of picoseconds width, which is also a typical signature of NLP laser.

A fast photodetector together with an oscilloscope are used to measure the output laser pulse train from the laser cavity (Fig. 2(j-n)). Fig. 2(n) illustrates the characteristic pulse train when the laser operates at 1712 nm, which shows some intensity fluctuations from pulse to pulse. This can be understood in the following way. The pulse bunches circulating in the laser cavity consist of many sub-pulses with random intensity. During each round trip, the sub-pulses evolve independently driven by the inner cavity effects, including saturable absorption, reverse saturable absorption, dissipative processes, dispersive effects, and nonlinear effects. The result is that the output pulse bunches are different from round trip to round trip with different arrangements of sub-pulses.

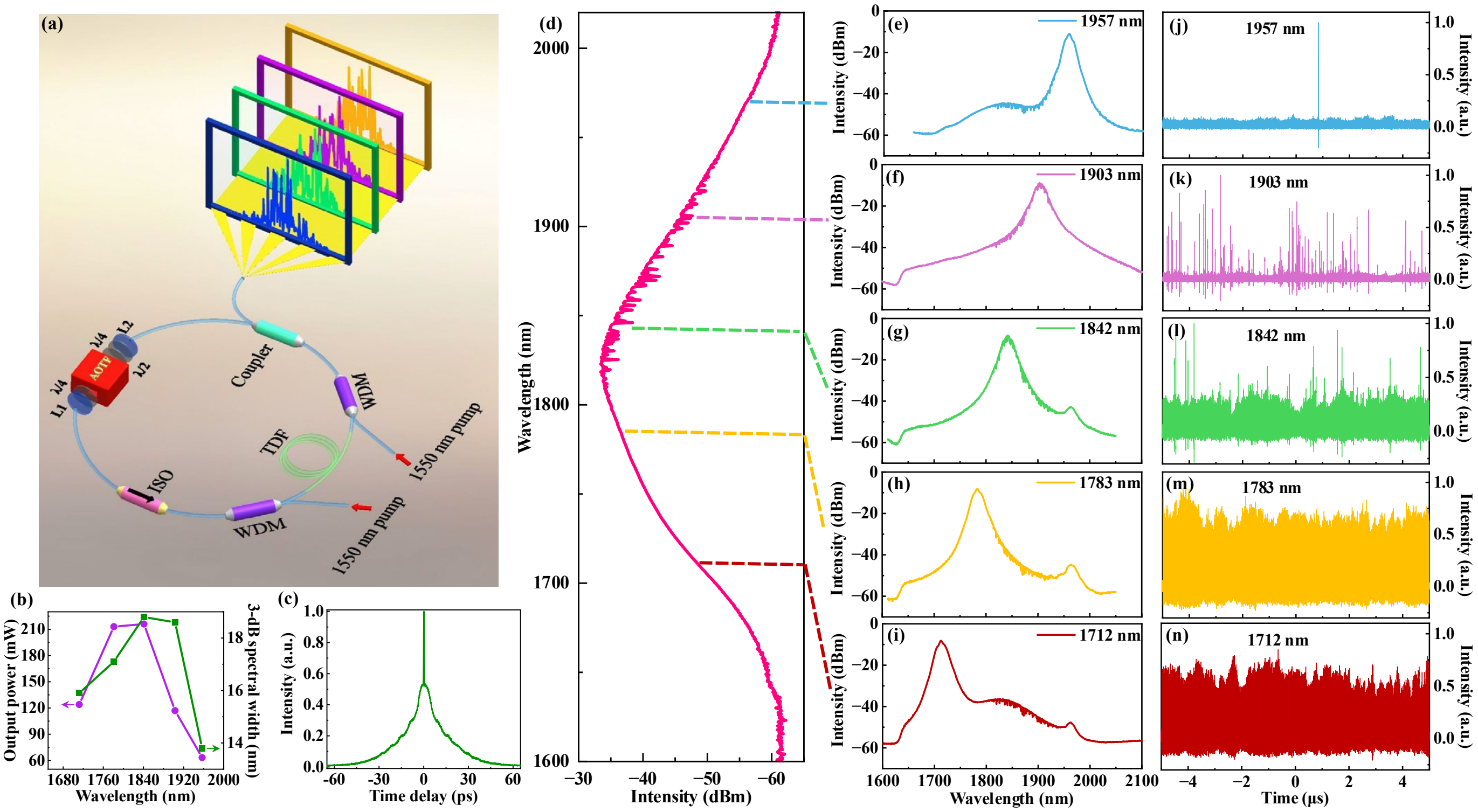

Fig. 2. (a) Experimental setup of the NLP laser; (b) output powers and 3-dB spectral widths at different wavelengths; (c) autocorrelation trace when laser operates at 1712 nm; (d) the amplified spontaneous emission spectrum of the TDF; (e-i) Measured output spectra of the tunable NLP laser with corresponding pulse trains (j-n) at 1957 nm, 1903 nm, 1842 nm, 1783 nm and 1712 nm central wavelengths.

As shown in Fig. 2(d), an experimentally measured amplified spontaneous emission from the TDF exhibits the water absorption lines within the laser operation range. When the laser operation wavelength is tuned from 1712 nm into the water absorption spectral region, the pulse train recorded by the oscilloscope shows some random giant pulses (Fig. 2 (j-l)). However, this phenomenon is not observed when the Tm-doped laser operates in the same wavelength region in conventional soliton and dissipative soliton regimes[16, 21]. We attribute this unique phenomenon of the NLP laser to a rare statistical event, the generation of rogue waves[22]. The origin is attributed to the unique gain/loss mechanism inside the laser cavity. The single-shot spectra of NLP pulses contain many sub-pulses[23]. As the pulses evolve in the laser cavity, the pulse bunch experiences a combined gain/loss mechanism, including gain provided by the active fiber, loss introduced by the water absorption, and other intra-cavity losses. In rare cases, the spectrum of the pulse bunch can evolve in a manner that avoids the water absorption lines, resulting in a pulse bunch with significantly higher energy.

**Pulse dynamics and operational regimes of the NLP laser: simulation and experiment**

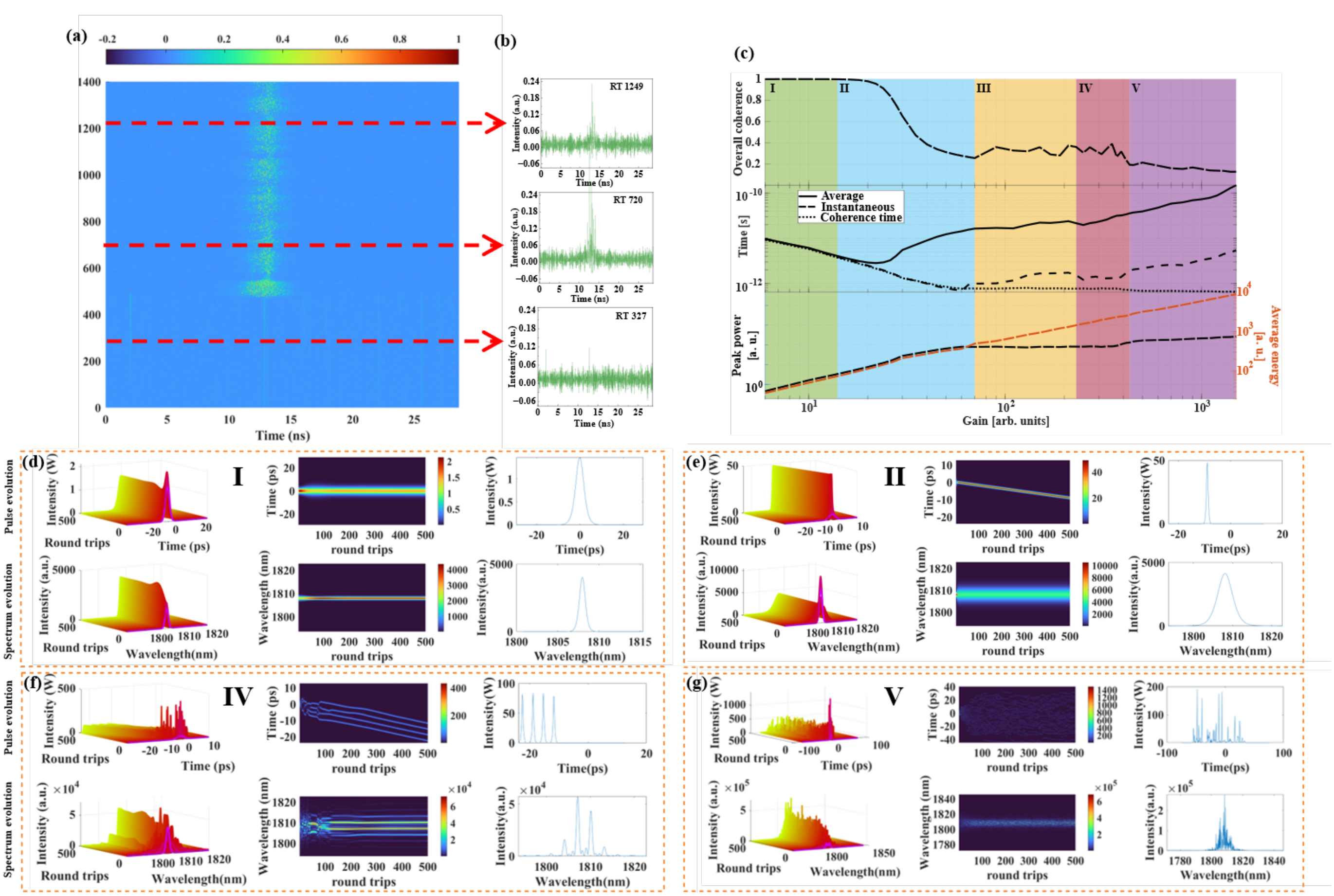

Fig. 3. Pulse evolution dynamics. (a) Experimentally measured NLP buildup process by DFT method; (b) corresponding instantons oscilloscope traces of RT=1249 RT =720 and RT=327; (c) Numerical analysis of pulse evolution dynamics depending on the gain parameters; Different operation regimes: coherent regime (I), drifting regime (II), splitting regime (III), transient regime (IV), and chaotic regime (V). Upper panel: Overall degree of coherence; middle panel: averaged pulse burst width (solid line), averaged pulse width of the sub-pulses (dashed

line), and coherence time (dotted line); Lower panel: averaged peak power (black long dashed) and average pulse energy (red long dashed). (d-g) Simulated laser starting dynamics of different operation regimes at small-signal gain values of 10 $m^{-1}$ (I), 40 $m^{-1}$ (II), 230 $m^{-1}$ (IV) and 1000 $m^{-1}$ (V). Left panel and middle panel: roundtrip-to-roundtrip pulse and spectrum evolution in 3D and 2D plotting format, respectively; Right panel: the pulse intensity and spectrum profiles at RT=500.

To reveal the evolution of pulses dynamics in the NLP laser, we probe the buildup process by measuring single-shot spectra with the DFT[24] method and carrying out numerical simulations at different gain levels following the experimental parameters, with a focus on the evolution of coherence properties (see Methods for details). Fig. 3 (a) illustrates experimental observation of roundtrip-to-roundtrip pulse bunch evolution around the starting point of NLP operation. Before the starting point, several weak pulses exist in the cavity. Since the laser utilizes NPR as the mode-locking technique, the weak pulses adjust their polarization states in every roundtrip (RT), until around RT=500 the positive feedback from the laser cavity supports laser pulse formation from one weak pulse. Together with soliton collapsing[25], the pulse bunches form and maintain. At the same time, other weak pulses all die out due to gain depletion. The instantaneous oscilloscope traces of RT=327, RT=720 and RT=1249 are shown in the right panel of Fig. 3(a), clearly illustrating the transition between operation regimes.

By conducting numerical simulations, we observe that the laser cavity operates in different regimes once applying different gain levels (see the Materials and Methods section for more details of simulation configuration). The simulation results are summarized in Fig. 3(c-g). The simulated intracavity field converges to either a single pulse, multiple pulses, or a noise-like pulse with the increase of small signal gain. The single pulse regime exists in the small signal gain range from 6 $m^{-1}$ to 70 $m^{-1}$ (Fig. 3 (d, e)). With the increase of gain, the simulated pulse durations of the converged pulse decrease from 7.7 ps to 293 fs. The spectral bandwidth is enhanced from 0.4 nm to 5.67 nm, and the pulse begins to drift in the temporal domain at higher gain levels. When the small signal gain is larger than 90 $m^{-1}$, the laser cavity begins to support multiple pulses; for example, at a small signal gain of 230 $m^{-1}$, the laser cavity contains 4 pulses (Fig. 3(f)). When the small signal gain is larger than 430 $m^{-1}$, the laser pulse begins evolving into a chaotic state, which is the noise-like pulse regime (Fig. 3(g)).

**Coherence analysis**

By using the simulated and measured results of pulse dynamics evolution, we evaluate the coherence time of NLPs. The earlier attempt to measure coherence time was challenged by insufficient resolution of the measurement system, revealing that the coherence time of NLPs is much smaller than a conventional mode-locked laser[15]. To overcome this obstacle, we exploit two distinct approaches carried independently of each other supplemented with detailed simulations. This enables us to provide a realistic estimation of temporal coherence. The first method is based on the single-shot spectra measured via DFT, which are qualitatively compared to the numerically obtained single-shot spectra. Through the comparison, we are able to establish that the variation in the spectrum is likely caused by partial coherence, which allows us to construct the second-order correlation function without knowledge of the spectral phase (see Methods for details). The second method relies on multi-pulse measurements carried out with frequency-resolved optical gating. It is well-known that the failure of the frequency-resolved optical gating (FROG) algorithm to converge is directly linked to low temporal coherence[26,27]. After confirming that the algorithm indeed does not converge, we employ a modified FROG algorithm where the measured single-shot spectra are used as additional retrieval constraints. This allows us to retrieve a realistic estimation of the pulse train and thus form the relevant correlation functions.

For coherence length estimation from numerical simulations, we model 1000 roundtrips inside the laser cavity for each gain value. The coherence properties of the resulting pulses are evaluated by calculating the mutual coherence function (MCF) and cross-spectral density (CSD). The calculation of the two quantities is elaborated in the Materials and Methods section. Since the pulse is coupled out from the cavity at every round trip, the simulated pulses essentially form a pulse train, which forms the actual pulse train emanating from the laser. We note that transient effects usually occur up to 200 round trips from the beginning of the operation, which may skew the coherence estimation. To ensure that the intracavity pulse has converged to a stable solution (if one exists), we used only the last 500 realizations for the MCF. We then proceeded to compute the time-averaged MCF and the overall degree of coherence. We further estimated the coherence time with two different methods, one based on the Fourier transform of the pulses and the other on the width of the time-averaged MCF, finding good agreement between them. Additionally, we computed the trends for average peak power and average pulse energy. The pulse properties are presented in Fig. 3(c) as a function of gain.

From Fig. 3 (c), we can identify five regions where the dynamics of the laser are markedly different. First, in region I, the cavity contains a single highly coherent pulse (a soliton solution). This is evident from the overall degree of coherence as well as the coinciding pulse durations and coherence times (i.e. the pulse is bandlimited in width). Therefore, we call this the coherent regime. The pulse is on the order of picoseconds in FWHM. As the gain increases, the pulse becomes narrower, and its peak power increases rapidly. In this region, the increase in the average energy of the pulse correlates with the increase in peak power.

After reaching a certain threshold gain, the pulse begins to experience a timing drift, which is associated with a dramatic decrease in the overall degree of coherence in the drifting region II. That is, the pulse moves within the sampling window, which decreases the overall degree of coherence, although the pulse remains essentially bandlimited. The effect of the drift is analogous to timing jitter, and while it does not affect the shape or the width of the instantaneous intensity, it can greatly increase the length of the *average* pulse. This is what causes the decrease in the overall degree of coherence, although the instantaneous intensity still coincides with the coherence time. The increase in the average energy of the pulse again correlates with the increase in peak power.

As we increase the gain further, the amplitude of the pulse increases enough to cause soliton fission. In other words, the pulse has gained enough energy to split into multiple fundamental solitons, with the number of sub-pulses depending on the amount of gain. This is the splitting region III. We now note that the overall degree of coherence slightly increases at first and then starts to decrease again. Whenever a pulse splits, it also experiences less timing drift, which results in increased coherence. Additionally, the instantaneous pulse width is now slightly higher than coherence time, and thus the sub-pulses are longer than their bandlimit. Note that the width of the entire multi-pulse solution is much larger, and this width corresponds to the average width of the individual peaks, which make up the total solution. Further, the peak power and average energy no longer increase at the same rate. In fact, the peak power plateaus within region III, while average energy continues to increase exponentially. This is a desirable property for bioimaging since a high peak power is more damaging than a large pulse energy, and both increase brightness.

Region IV is the transient regime, where the field appears to present a chaotic time evolution but settles into a well-behaved multi-pulse solution after sufficiently many round trips. Here, the overall degree of coherence oscillates unpredictably, and the width of the peaks in the instantaneous intensity distribution becomes slightly narrower (Fig. 3 (c), middle panel, dashed line). Moreover, the spectrum features strong shot-to-shot variation until the field settles to a well-behaved solution. After converging, the spectrum is constant.

At the highest gain level lies region V, which is the chaotic region. When the cavity is saturated with electromagnetic energy, the time-evolution is chaotic: pulses merge, split, and drift in unpredictable ways, which is directly reflected in the noisy spectral evolution. That is, the spectrum varies from shot-to-shot, and the NLP laser operates in this regime. Region V is marked by the near constant decrease of the overall degree of coherence, an increase in the instantaneous pulse width, as well as a small jump in the average peak power of the pulses.

Ideally, one would measure the shape of individual pulses to detect whether the chaotic region has been reached. However, this was not possible in our case. There are two experimentally available avenues for detecting this region when measurements of the shape of single pulses are not possible. The first is to look at the spectra of individual pulses. While the average spectrum is relatively smooth, the pulse-to-pulse spectrum variation is very noisy. Another clear sign of the chaotic region is seen when one attempts to measure the average pulse shape with frequency-resolved optical gating. If the retrieval algorithm does not converge, and the FROG error remains high, then the temporal coherence of the pulse train is low[26], since the measured trace is an incoherent sum over a large number of coherent FROG traces[27].

With the use of DFT, we were able to measure a set of single-shot spectra, which are fairly representative of the undisturbed single-shot spectra. A small subset of the measured spectra is presented in Fig. 4a, together with the measured FROG interferogram in Fig. 4b. From the measured spectra, we can see that they feature large shot-to-shot variation, which is already a tell-tale sign of the chaotic region. Moreover, the FROG interferogram has a sharp peak in the middle, with a broad background distribution. This is also a well-known sign of low coherence, and the width of the sharp peak is comparable to coherence time, whereas the broad background has a width comparable to the average pulse length.

The result of spectral coherence computation from measured single-shot spectra is shown in Fig. 4c, see Methods section for details. This is a good indicator of low coherence, but it relies on the assumption of Gaussian statistics, and it is quite noisy. Therefore, we aim to quantify the correlations with a different method.

We probe the spectral and temporal coherence of the NLP laser by employing FROG measurements over an average of one million pulses. The FROG retrieval is performed with a specialized algorithm, which imposes the measured spectra onto the retrieved field on every iteration cycle as an additional constraint. This way, we can constrain the retrieval in both time (via FROG trace) and frequency. The FROG error is quantified with the usual rms-approach, and it never went below 0.02 during the retrieval process. Moreover, the retrieval algorithm does not converge on a single solution. The additional measurement constraint contracts the space of possible solutions, allowing us to estimate the pulse properties more reliably. We then pick solutions with a low FROG error as representative of the actual pulse train and repeat the process for several measured spectra. The retrieved pulses are then used to compute the MCF and CSD, and to estimate the coherence properties of the laser. The resulting correlation functions are shown in Fig. 4d.

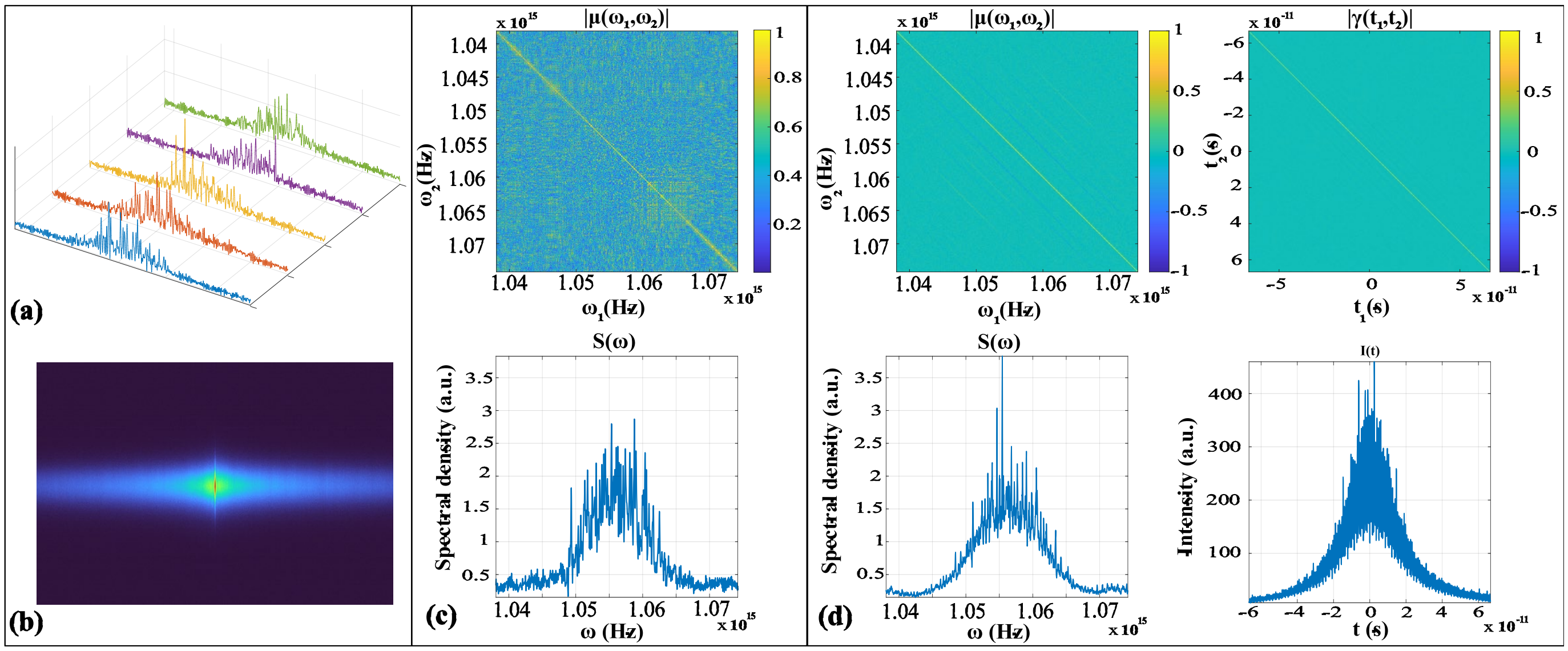

Fig. 4. Coherence analysis of the noise-like pulsed laser. (a) Samples of the measured single shot spectra; (b) the measured FROG spectrogram;(c) absolute value of the spectral degree of coherence evaluated from the measured single shot spectra (upper panel), and the corresponding average spectrum (lower panel), (d) results from the FROG retrieval. Left upper panel: the spectral degree of coherence; left lower panel: the corresponding average spectrum; right upper panel: the temporal degree of coherence; right lower panel: the corresponding average pulse shape.

From Fig. 4d one can see that the degree of spectral coherence obtained this way has a very similar structure as the one estimated from spectrum measurements in Fig. 4c. Notably, the NLP laser is estimated to be very incoherent, with a coherence time of ~100 fs (width of the anti-diagonal of the temporal correlation function) and an average pulse length of ~40 ps (width of average intensity). Note that since we could not directly measure the individual pulse shapes and instead indirectly estimated them, the constructed correlation functions are not exact. However, due to the strong constraints and the use of two complementary approaches supplemented with numerical simulations ensure that the estimated correlation functions represent the properties of these particular pulse trains fairly well. We further validated this by reconstructing the measured FROG trace with the pulses picked for the correlation functions. They reproduced the trace with sufficient accuracy.

Finally, we compared the coherence time from the MCF with the coherence time one would expect from the single-shot spectra. The single-shot spectra featured coherence times on the order of 140-160 fs, with an average of 156 fs, which corresponds well with the approximated degree of coherence. Note that the coherence times estimated from single-shot spectra are expected to be different since one is not averaging over many pulses. Thus, it is safe to conclude that the NLP laser produces light with a low coherence, low peak power, high brightness, wide spectrum, and can be widely tunable pulses over several hundred nanometers.

**Discussion**

The rapid advancement of light-driven imaging modalities continues to impose increasingly stringent requirements on light sources. Imaging performance, such as particularly contrast, resolution, and penetration depth, is governed by laser parameters including spectral bandwidth, coherence properties, and brightness. However, optimizing these characteristics simultaneously within a single source remains challenging due to inherent trade-offs. Among these, wide spectral tunability and low coherence are especially difficult to achieve, as each introduces distinct constraints on laser design and performance.

Tunability is particularly important for applications requiring wavelength scanning, where flexible control over emission wavelength and bandwidth is essential. Frequency sweeping can generally be realized in two ways: (i) indirectly, using nonlinear parametric processes, or (ii) directly, using wavelength-selective components. Indirect methods often require complex setups with separate tuning stages, while direct approaches allow integration of tuning elements into the laser cavity, enabling more compact and efficient systems. An acousto-optic tunable filter (AOTF) provides an effective solution for intracavity wavelength selection. Its operation relies on the phase-matching condition of acousto-optical diffraction, enabling precise and controllable wavelength tuning[28]. The transmission band of commonly used acousto-optic crystals, such as $TiO_2$, extends into the mid-infrared region, allowing operation over a broad spectral range[29]. Compared with Lyot filters, which can also achieve wavelength tuning, an AOTF transmits a single dominant wavelength, which is critical for isolating specific spectral components. Additionally, because it is driven by a radio-frequency signal generator, it enables rapid spectral tuning and sweeping. These features make AOTFs highly suitable for constructing wavelength-tunable pulsed laser sources with external control[30]. In this work, we demonstrate a wavelength-tunable laser operating in the NLP regime across the full gain spectrum (1650–2000 nm) of a TDF laser using an intracavity AOTF. To the best of our knowledge, this represents the widest tunability reported for NLP lasers. This broad spectral coverage offers significant advantages for imaging applications.

In particular, the demonstrated tuning range is highly relevant for imaging within the third biological window (~1650–1870 nm), where reduced optical scattering enables deeper tissue penetration[18]. Coverage from 1650 nm to 2000 nm allows flexible selection of excitation wavelengths for different fluorophores and nonlinear processes. It also supports multi-contrast imaging, spectrally resolved nonlinear microscopy, and OCT. Extending toward longer wavelengths can further enhance imaging contrast and versatility.

Low coherence is another critical requirement in advanced imaging systems. It is typically associated with broadband, temporally fluctuating emission, which complicates the maintenance of both stability and brightness. Coherence length is the path length difference over which light can produce visible interference fringes when it is interfering with itself[31]. As such, it is the main quantity of interest for interferometric techniques such as OCT due to the direct effect on axial resolution or multiphoton imaging for reduction of speckles. While coherence time is often used as a descriptor, it depends only on the spectral bandwidth and does not fully capture field correlations. As a result, two sources with identical spectra exhibit exactly the same coherence lengths, even if their overall degrees of coherence are very different. Evaluating the actual degree of coherence is therefore essential, as it directly affects peak power. Mode-locked femtosecond lasers can provide comparable spectral bandwidths but exhibit much higher peak powers due to their high coherence, which may lead to photodamage in biological samples. In contrast, NLP operation produces a fundamentally different temporal structure: the output consists of many randomly distributed sub-pulses grouped into bursts of approximately 40 ps duration. This structure distributes energy over multiple events, significantly reducing peak intensity while maintaining useful average power.

The effects of such pulse bursts differ between linear and nonlinear imaging modalities. In linear techniques like OCT, the detected signal depends primarily on average power and coherence length, allowing multiple low-peak-power pulses to accumulate and improve the signal-to-noise ratio. In nonlinear imaging methods, signal generation depends on peak intensity; however, if individual sub-pulses have sufficient intensity, nonlinear processes can still occur efficiently. Additionally, the low second-order coherence of NLP lasers may reduce interference and speckle artifacts in scattering media. Although the demonstrated source does not yet provide sufficient peak power for nonlinear excitation, this limitation can be addressed through moderate amplification stages, as shown in previous studies[32,33].

**Material and Methods**

**DFT measurement**

To reveal the transient pulse dynamics in the NLP laser, we investigate the buildup process of the NLP laser using the DFT method[24], when the laser spectrum is centralized at the ~1783 nm wavelength. A spool of 4.8 km SMF28 fiber is spliced with a laser output port to stretch the pulse bunches, which are stretched to around 5 ns as shown in Fig. 3. The data acquisition is implemented by a fast photodetector (Newport 1414, 25 GHz bandwidth) and a data analyzer (Tektronix DSA 72004) with 20 GHz bandwidth and 50 GS/s sampling rate.

**Laser cavity simulation**

We employ numerical simulation in our analysis, as they include complete information of both field amplitude and phase. The parameters of the simulated laser cavity have been chosen to match the experimental one. The light propagation in optical fiber is modeled with the generalized nonlinear Schrödinger equation

$$\frac{\partial A}{\partial z} + i\frac{\beta_2}{2}\frac{\partial^2 A}{\partial T^2} = \frac{g}{2}A + i\gamma\left(1 + i\tau_{shock}\frac{\partial}{\partial T}\right)\left(A\int_{-\infty}^{\infty} R(t)\,|A(z, T-t)|^2 dt\right),$$

where $A(z, t)$ is the complex electric field envelope, $z$ is the coordinate of pulse propagation, $T$ is the retarded time coordinate moving at the group velocity of pulse, $\beta_2$ is the group velocity dispersion (GVD) parameter, $g$ is the gain coefficient of the active fiber, $\gamma$ is the nonlinear coefficient, $\tau_{shock}$ is the shock-formation time-scale that is responsible for the self-steeping effect, and $R(t)$ is the response function that models both the instantaneous electronic (Kerr) nonlinearity and the delayed molecular (Raman) nonlinearity[34]. The generalized nonlinear Schrödinger equation is solved with the split-step Fourier method using second-order Runge-Kutta Algorithm[35]. In the active fiber, the gain is modeled with the saturation model, $g = g_0/(1 + E_{pulse}/E_{sat})$, where $g_0$ is the small-signal gain coefficient, $E_{pulse}$ is the pulse energy, and $E_{sat}$ is the gain saturation energy. In the simulation, $\beta_2$ is set as -0.066 $ps^2/m$ for both SMF28 fiber and Tm-doped fiber. $\gamma$ is set as 0.0023 $W^{-1}m^{-1}$. To simulate the reverse saturable absorption effect that is essential for noise-like pulse generation, a sinusoidal model is adopted to model the effect of nonlinear polarization rotation[36]:

$$T' = T_0 + dT \cdot sin^2\left(\frac{\pi}{2}\cdot\frac{P}{P_A} + \varphi\right)$$

Where $T'$ is the transmission of the artificial saturable absorber, $T_0$ is set to 0.05, $dT$ to 0.95, $P_A$ to 250 W, and $\varphi$ to π/4. The AOTF is modeled with a Gaussian-shaped spectral filter with a bandwidth of 12 nm full-width at half-maximum. 30% loss is added after the spectral filter to simulate the coupling loss in a real experiment, and 20 % optical power is extracted out at the position of the output coupler.

**Correlation functions**

We estimate the coherence properties by constructing the MCF and CSD, defined as

$$\Gamma(t_1, t_2) = \langle E^*(t_1)E(t_2)\rangle,$$

and

$$W(\omega_1, \omega_2) = \langle E^*(\omega_1)E(\omega_2)\rangle,$$

respectively. Here, asterisk denotes complex conjugation, $t$ is time, $\omega$ is angular frequency, and the angle brackets stand for ensemble averaging. The electric field $E$ is taken in the moving reference frame of the pulse[37]. The MCF contains the average pulse intensity $I(t) = \Gamma(t,t)$, whereas the CSD contains the average spectrum $S(\omega) = \mathrm{W}(\omega,\omega)$. One can use these to normalize and obtain the complex degree of coherence in both temporal and spectral domains

$$\gamma(t_1,t_2) = \frac{\Gamma(t_1,t_2)}{\sqrt{I(t_1)I(t_2)}},$$

$$\mu(\omega_1,\omega_2) = \frac{\mathrm{W}(\omega_1,\omega_2)}{\sqrt{S(\omega_1)S(\omega_2)}}.$$

To obtain a single numerical value that represents the coherence of the field, we can compute the overall degree of coherence as in

$$\bar{\gamma}^2 = \frac{\iint |\Gamma(t_1,t_2)|^2 dt_1 dt_2}{\iint I(t_1)I(t_2) dt_1 dt_2},$$

which has the same value in both domains.

In addition to the degree of coherence, we wish to find the coherence time as well. There are essentially two ways to find coherence time, $\tau$. The first one is to compute the time-averaged MCF by integrating over average time $t = (t_1 + t_2)/2$ and finding the $1/e$ width of the resulting distribution. The second is to take the ensemble averaged pulse spectrum, find the full-width at half-maximum (FWHM), and use it to compute the coherence time via $\tau = 1/\pi\Delta\nu$, where $\Delta\nu$ is the FWHM in frequency. As stated above, coherence time is important because it tells us the path length difference over which visible interference fringes can be observed. Additionally, it gives us the temporal width of the shortest possible pulse for a given spectrum. We save the $1/e$ width of the average pulse intensity $I(t) = \Gamma(t,t)$, as well as the $1/e$ width of the peaks in the instantaneous intensity distribution $|E(t)|^2$, for comparison with the coherence time.

**Spectral coherence estimation from measured single-shot spectra**

The spectral coherence of the source is estimated with the measured spectra, by constructing the spectral energy correlation function $W^{(e)}(\omega_1,\omega_2) = \langle S_i(\omega_1)S_i(\omega_2)\rangle$, where $S_i(\omega)$ are single-shot spectra. This function is a special case of the fourth order correlation function $W^{(4)}(\omega_1,\omega_2,\omega_3,\omega_4) = \langle E^*(\omega_1)E^*(\omega_2)E(\omega_3)E(\omega_4)\rangle$, which is obtained by setting $\omega_3 = \omega_1$ and $\omega_4 = \omega_2$. If we assume that the correlation function is Gaussian and employ Isserlis' theorem, as in

$$W^{(4)}(\omega_1,\omega_2,\omega_3,\omega_4) = \langle E^*(\omega_1)E^*(\omega_2)\rangle\langle E(\omega_3)E(\omega_4)\rangle + \langle E^*(\omega_1)E(\omega_3)\rangle\langle E^*(\omega_2)E(\omega_4)\rangle + \langle E^*(\omega_1)E(\omega_4)\rangle\langle E^*(\omega_2)E(\omega_3)\rangle$$

we can then set $\omega_3 = \omega_1$ and $\omega_4 = \omega_2$, to obtain

$$\begin{aligned} W^{(4)}(\omega_1,\omega_2,\omega_1,\omega_2) &= \langle S_i(\omega_1)S_i(\omega_2)\rangle \\ &= \langle E^*(\omega_1)E^*(\omega_2)\rangle\langle E(\omega_1)E(\omega_2)\rangle + \langle E^*(\omega_1)E(\omega_1)\rangle\langle E^*(\omega_2)E(\omega_2)\rangle + \langle E^*(\omega_1)E(\omega_2)\rangle\langle E^*(\omega_2)E(\omega_1)\rangle \\ &= S(\omega_1)S(\omega_2) + |W(\omega_1,\omega_2)|^2. \end{aligned}$$

Therefore, it is possible to write

$$|W(\omega_1,\omega_2)| = \sqrt{\langle S_i(\omega_1)S_i(\omega_2)\rangle - S(\omega_1)S(\omega_2)}$$

which can be normalized as usual to obtain the absolute value of the complex degree of coherence.

**Acknowledgments**

This work was supported by the Research Council of Finland (346518, 346511, 320165).